# Metal Object Detection Based on Load Impedance and Input Power Characteristics for High-dimensional WPT System

Zixi Liu, Qi Zhu, and Mei Su

*Abstract*—High-dimensional wireless power transfer (WPT) systems have received increasing attention for charging mobile devices. With the receivers have higher spatial freedom, the systems are more susceptible to the metal objects in the surroundings. However, conventional methods for metal object detection (MOD) can't satisfy the requirements of safety and stability of new systems. This paper proposed a metal detection method which is more suitable for high-dimensional WPT systems. The key feature of the proposed method is combining load impedance and input power characteristics curves to determine whether the receiver is a metal object or a coil. And the influence of the metal is discussed in the circuit and magnetic model. The method is theoretically proven by the simulation calculation. And the experiments verified that the method can accurately detect the metal objects by setting the threshold curves.

## I. INTRODUCTION

As an emerging transfer technology, wireless power transfer (WPT) has attracted much more attention than before, after the theory of coupled magnetic resonance has been proposed [1]. The system efficiency and the transfer distance had been improved significantly via magnetic resonance. Therefore, the WPT system is being widely used in electric vehicles, mobile devices and other industrial and home applications [2-12]. However, the energy in the WPT system is transferred by the coupled magnetic field. Since the metallic or magnetic material objects can absorb energy from the magnetic field in the form of heat and become a hazard [4], the detection of metallic or magnetic material objects around the system is crucial to the safe operation. In order to increase the safety and reliability of the WPT system, several different detection methods have been proposed in previous studies [2-7]. Detection methods can be divided into two categories, one is analyzing the existing parameters of the system [2-6], and the other is adding extra detecting devices [7].

It is observed that there is a significant decrease in quality factor Q of the receiving coil when there is a metal object between the transmitter coil and the receiving coil. S Fukuda et al in [2] designed a circuit for Q value measurement to detect the presence of the metal object. By analyzing the equivalent circuit of the WPT system, [3] found that resonant frequency and the transmission efficiency varied a lot when the metal object enters. H Kudo et al in [4] proposed a detection procedure by detecting the frequency bifurcation and comparing peak frequency. N Kuyvenhoven et al in [5] studied the relationship between the eddy current loss and the current in coils and pointed out that evaluating the power loss of the system can identify the foreign object in real time. Z N Low et al in [6] designed a set of sampling circuits to analyze the transmitter coil voltage and supply current, and choose the appropriate working mode for the system according to the sampling results. Meanwhile, a non-overlapping coils detection circuit is proposed in [7], this additional device can detect not only the presence of the metal object but also the location of the metal. However, this detection circuit is only suitable for flat charging system like electric vehicles.

All the detection methods mentioned above are based on the planar charging system. The transmitter is usually placed on the ground or a desk, hence only the space between the transmitter and the receiver needs metal detection. The size of this space only depends on the size of the coils and the distance between the transmitter and the receiver. Those detection methods may not fit spatial WPT systems with the large freedom of movement of power pickups.

P Raval et al in [9] and [10] developed a 3D inductive power transfer system for industrial battery charging. To improve the transfer distance and efficiency, the technique of shaping the magnetic field in a high-dimensional WPT system is proposed in [11]. [11] designed a novel WPT system in which the shaped magnetic field forms an energy beam that directs to the receiver. As a result, the leakage magnetic field is reduced significantly compared to the conventional inductive WPT system. For more intelligent charging applications, [12] and [13] studied an omnidirectional WPT system and summed up the basic control principles. Based on the information about the voltage and current in the transmitter circuit, [14] and [15] proposed an active filed orientation method to shape the magnetic flux and realized three-dimensional full-range field orientation. Most of these novel systems are used for home applications. The charging range of those systems extends from the plane space between two coils to the whole space around a set of orthogonal transmitter coils. As a result, the system is more susceptible to the metal which is commonly used at home.

This paper aims to solve the problem of MOD in high-dimensional WPT system. By analyzing the impedance and the power characteristics curves of the metal object and the coil, a detection method is proposed without communication between the transmitter side and the receiver side. The major study in this paper can be summarized as follows: 1) a precise circuit and magnetic combined model of a high-dimensional WPT system is established; 2) the load impedance and input power characteristics under different conditions are analyzed and a MOD method is proposed; 3) a prototype of WPT system with the function of MOD is implemented.

The rest of this paper is organized as follows: In Section II, the high-dimensional wireless power transfer system is introduced based on magnetic and circuit analysis. In Section III, the different influence of receiver coil and metal are discussed through the simplified equivalent model, and a method for MOD is proposed based on the load impedance and input power characteristics. In Section IV, the hardware implementation for MOD is introduced, and the feasibility of the proposed detection method is verified experimentally. In Section V, a conclusion is drawn, and future studies based on proposed methods are suggested.

## II. THEORETICAL ANALYSIS OF HIGH-DIMENSIONAL WIRELESS POWER TRANSFER SYSTEM

### A. Structure of High-Dimensional WPT System

A high dimensional wireless power transfer system is shown in Fig. 1. The transmitter coils used in the WPT system are in the shape of a square with their center overlapped, and the plane of each coil is perpendicular. Two coils marked as coil A and coil B respectively, are orthogonal transmitter coils. Coil A is on the XOZ plane, while coil B is on the YOZ plane. Each coil connected with an AC voltage source, a series resonant capacitor. Coil C is a receiver coil loaded with a series resonant capacitor and a resistive load.

The receiver coil can be positioned in any space around the transmitter coils, and the main magnetic flux is controlled pointing the center of the receiver.

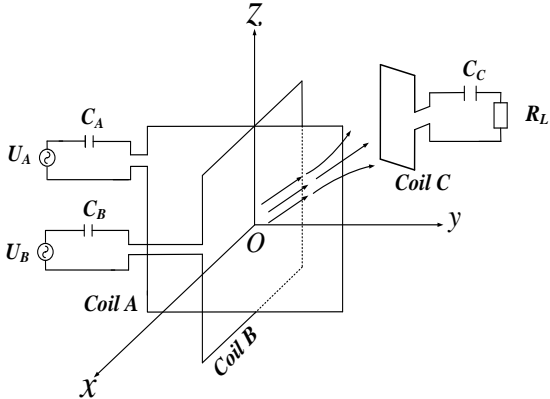

Fig. 1 A high dimensional WPT system.

### B. Magnetic Field Analysis

Fig. 2 is the vertical view of the proposed WPT system, and it describes the geometric relationship between the transmitter coils and the receiver coil.

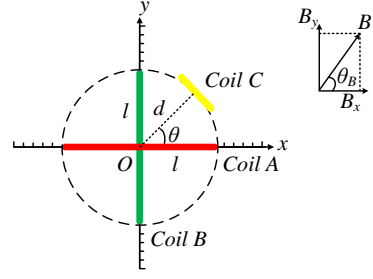

Fig. 2 Geometrical relationship between the transmitters and the receiver coils in XOY plane.

In Fig. 2, coil A and coil B are on the X-axis and Y-axis, respectively. The length of the transmitter coils is $2l$. The receiver coil C is placed around coil A and coil B with its plane facing the center of the transmitter coils, and the distance between the center of transmitter coils and the center of receiver coil is $d$. The line connecting the center of coil C and the origin is $\vec{d}$, and the angle between $\vec{d}$ and the X-axis is $\theta$. It is obvious that the maximal coupling between the transmitter coils and the receiver coil can be achieved when the main shaped magnetic field vector $\boldsymbol{B}$ is perpendicularly pointing to the center of coil C. The angle between $\boldsymbol{B_x}$ and $\boldsymbol{B}$ is assumed as $\theta_B$. $\theta$ will be equaled to $\theta_B$ when the direction of the total magnetic flux density $\boldsymbol{B}$ is pointing towards coil C. From the analysis above, the relationship between the total magnetic flux density $\boldsymbol{B}$ and the magnetic flux density generated by each transmitter coil $\boldsymbol{B_x}$ and $\boldsymbol{B_y}$ can be expressed follows.

$$\begin{cases} B_x = B\cos\theta \\ B_y = B\sin\theta \end{cases}, (0 \leq \theta \leq 2\pi) \quad (1)$$

In order to figure out the value of $\theta$, equation (1) also can be expressed as follows.

$$\begin{cases} |B| = \sqrt{B_x^2 + B_y^2} \\ \theta = tan^{-1}\left(\frac{B_y}{B_x}\right) \end{cases}, (0 \leq \theta \leq 2\pi) \quad (2)$$

From equation (2), it can be known that the total magnetic flux density $\boldsymbol{B}$ can be pointed to any angle on a plane by controlling the magnitude and direction of $\boldsymbol{B_x}$ and $\boldsymbol{B_y}$. In order to simplify the analysis of the magnetic flux density, including the magnitude and direction, taking the origin of coordinates $O(0,0,0)$ as an example to calculate $\boldsymbol{B_x}$ and $\boldsymbol{B_y}$. Assumed two currents on coil A and B are $I_A\cos(\omega t + \varphi_A)$ and $I_B\cos(\omega t + \varphi_B)$, respectively. Each axis component of $\boldsymbol{B}$-field at (0,0,0) is shown as follows.

$$\begin{cases} B_{x0} = -\frac{\sqrt{2}\mu_0 N I_B \cos(\omega t + \varphi_B)}{\pi l} \\ B_{y0} = -\frac{\sqrt{2}\mu_0 N I_A \cos(\omega t + \varphi_A)}{\pi l} \end{cases} \quad (3)$$

where $\omega$ is the angular frequency of two independent voltage sources. $I_A$, $I_B$ are the amplitudes of the currents in the transmitter coils, respectively. $\varphi_A, \varphi_B$ are the phase angles of two currents, respectively. $\mu_0$ is permeability of vacuum, valued $4\pi \times 10^{-7} \, N \cdot A^{-2}$.

Assumed $\varphi_B = 0$, $\varphi_A - \varphi_B = \triangle \varphi$, the expression of $\theta$ can be rewritten as follows.

$$\theta = tan^{-1}\frac{I_A\cos(\omega t + \Delta\varphi)}{I_B\cos(\omega t)} \quad (4)$$

In equation (4), $I_A$ and $I_B$ are controlled by the independent voltage source $U_A$ and $U_B$, respectively. If $\varphi_A = \varphi_B$, the phase of $I_A$ and $I_B$ will be the same phase. Whereas, if $\varphi_A = 180° - \varphi_B$, $I_A$ and $I_B$ will be the inverse phase. Regard $I_A$ and $I_B$ as two vectors, which only have two directions (same or inverse). And assume that the range of its amplitude is from zero to a constant number. In summary, a set of parameters (including $I_A$, $I_B$, $\varphi_A$ and $\varphi_B$) can be solved which make $\theta$ be any angle in the plane.

## C. Circuit Analysis

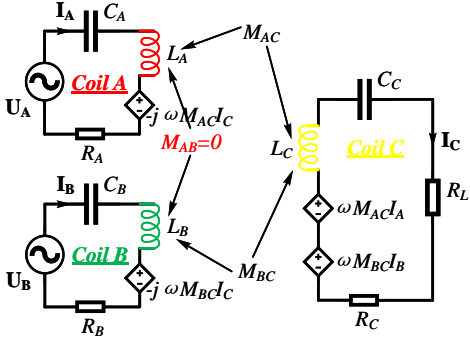

Fig. 3 Equivalent circuit model of the high dimensional WPT system.

The equivalent circuit of the system is constructed as Fig. 3 shows. $U_A$ and $U_B$ are excitation sources in the transmitter coils. $C_A$, $C_B$ and $C_C$ are the resonant capacitor of each transmitter coil and the receiver coil, respectively. $L_A$, $L_B$ and $L_C$ are the inductance of coil A, coil B and coil C, respectively. $R_A$, $R_B$ and $R_C$ represent the resistance of coil A, coil B and coil C. $I_A$ and $I_B$ are currents in the transmitter coils. $I_C$ is the load current.

Considering the transmitter coils are put in perpendicular, the mutual inductance between coil A and coil B is neglected. Based on Kirchhoff Voltage Law (KVL), the coupled circuit equivalent equations are given in (5).

$$\begin{cases} U_A - (-j\omega M_{AC}I_C) = (R_A + j\omega L_A + 1/j\omega C_A)I_A \\ U_B - (-j\omega M_{BC}I_C) = (R_B + j\omega L_B + 1/j\omega C_B)I_B \\ j\omega M_{AC}I_A + j\omega M_{BC}I_B = (R_C + R_L + j\omega L_C + 1/j\omega C_C)I_C \end{cases} \quad (5)$$

The mutual inductance between the transmitter coils and the receiver coil is decided by the coil shape and position. Once the position of the receiver coil is certain, the value of $M_{AC}$ and $M_{BC}$ will be certain. Adopting simplified reactance expression.

$$j\omega L_A + \frac{1}{j\omega C_A} = jX_A, j\omega L_B + \frac{1}{j\omega C_B} = jX_B, j\omega L_C + \frac{1}{j\omega C_C} = jX_C \quad (6)$$

where

$$\omega^2 = 1/L_AC_A = 1/L_BC_B = 1/L_CC_C$$

Considering the amplitude of total magnetic flux density **B** is constant and the direction of **B** can be any angle in the plane. That means $\theta \in [0, 2\pi)$. The relationship between $I_A$ and $I_B$ is summarized as follows [16].

$$\begin{cases} I_A = I\sin\theta \\ I_B = I\cos\theta \\ I = \sqrt{I_A^2 + I_B^2} \end{cases} \quad (7)$$

where $I$ is a sinusoidal time function which decides the amplitude of **B**. Assumed the parameters of two transmitter coils are exactly same, $R = R_A = R_B$, $X = X_A = X_B$. And substituting (6) and (7) into (5) results in (8).

$$\begin{cases} U_A + j\omega M_{AC}I_C = (R + jX)I\sin\theta \\ U_B + j\omega M_{BC}I_C = (R + jX)I\cos\theta \\ j\omega M_{AC}I\sin\theta + j\omega M_{BC}I\cos\theta = (R_C + R_L + jX_C)I_C \end{cases} \quad (8)$$

From (8), the current in the receiver coil is calculated as follows.

$$I_C = \frac{j\omega I(M_{AC}\sin\theta + M_{BC}\cos\theta)}{R_C + R_L + jX_C} \quad (9)$$

Rewritten (9) results in (10).

$$I_C = \frac{j\omega I\sqrt{M_{AC}^2 + M_{BC}^2}}{R_C + R_L + jX_C}\sin\left(\theta + \tan^{-1}\frac{M_{BC}}{M_{AC}}\right) \quad (10)$$

The input power of the system is the sum of power losses in each coil and the load power in the receiver coil. Input power calculation can be expressed as follows.

$$P_{in} = I^2\left[R + \frac{\omega^2(M_{AC}^2 + M_{BC}^2)}{(R_C + R_L)^2 + X_C^2} \cdot \sin^2\left(\theta + \tan^{-1}\frac{M_{BC}}{M_{AC}}\right)(R_C + R_L)\right] \quad (11)$$

In addition, the voltage on the transmitter coils can be calculated as follows.

$$\begin{cases} U_A = I\left[(R + jX)\sin\theta + \frac{\omega^2 M_{AC}(M_{AC}\sin\theta + M_{BC}\cos\theta)}{R_C + R_L + jX_C}\right] \\ U_B = I\left[(R + jX)\cos\theta + \frac{\omega^2 M_{BC}(M_{AC}\sin\theta + M_{BC}\cos\theta)}{R_C + R_L + jX_C}\right] \end{cases} \quad (12)$$

It can be seen from the above equations, both the input power of the system and the voltage on the transmitter coil are related to the load $R_L$. If the receiver part gives no information about the load, the only way to identify the type of the receiver is to analyze the relationship between $P_{in}$, $U_A$ and $U_B$.

## III. DETECTION OF METAL FOREIGN OBJECTS WITHOUT COMMUNICATION

From the analysis in Section II, the orientation of the receiver can be found through the input power of the system. However, for the safety of energy transmission, it is not enough to find the receiver, and it also needs to determine whether it is metal. Usually, the receiver has no feedback to the transmitter. Therefore, the way to distinguish a normal receiver and a metal object is by finding the differences of the parameters in the transmitter parts when the receiver is different. In the proposed MOD method, the input power of the system and the current or the voltage of transmitter coils are analyzed in detail under different circumstances.

### A. Model Simplification

The voltage on the transmitter coils depends on the total impedance of the system. The location of the receiver changes the equivalent impedance of the receiver. Noting that after the receiver enters the charging area, the voltage of transmitter coils keeps constant unless the receiver moves again. In order to simplify the analysis of the parameters in transmitter coils, two orthogonal transmitter coils can be regarded as one coil which is coaxial to the receiver during the detecting process. The equivalent relationship between the original model and the simplified model is analyzed as follows.

The equivalent circuit model of the system with only one transmitter coil is shown in Fig. 4. The left part is the transmitter while the right part is the receiver. $U_i$ is an independent voltage source, providing coil current $I_1$ which generates the **B**-field. $R_1$ and $L_1$ are the resistance and inductance of the transmitter coil, respectively. $C_1$ is the resonant capacitor in the transmitter part. $R_2$ and $L_2$ are the resistance and inductance of the receiver coil, respectively. And $C_2$ is the resonant capacitor in the receiver part. $U_P$ and $U_S$ are mutual inductance voltage on the transmitter coil and the receiver coil, respectively. $U_S$ provides load current $I_2$ in the receiver coil. Considering $M_{12} = M_{21} = M$. Equivalent circuit equations are given in (13)-(14).

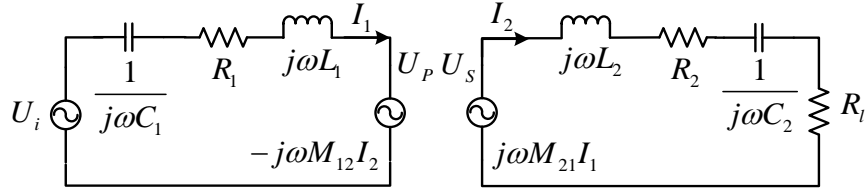

Fig. 4 Equivalent circuit of one transmitter coil system.

$$U_i + j\omega I_2 M = I_1(R_1 + jX_1) \quad (13)$$
$$j\omega I_1 M = I_2(R_2 + R_l + jX_2) \quad (14)$$

where

$$jX_1 = j\omega L_1 + \frac{1}{j\omega C_1}, jX_1 = j\omega L_1 + \frac{1}{j\omega C_1}$$

Considering the geometrical relationship between $\vec{B_x}, \vec{B_y}$ and $\vec{B}$, the formula $\vec{B} = \vec{B_x}\cos\theta + \vec{B_y}\sin\theta$ indicates the contribution made by each coil to the total magnetic flux density. Let the first two equations in (8) multiply $\sin\theta$ and $\cos\theta$, respectively, and adding up each other. (15) and (16) show the derivation results.

$$\sqrt{U_A^2 + U_B^2} \cdot \sin\left(\theta + \tan^{-1}\frac{U_B}{U_A}\right) + j\omega I_C(M_{BC}\cos\theta + M_{AC}\sin\theta) = I(R + jX) \quad (15)$$

$$j\omega I(M_{BC}\cos\theta + M_{AC}\sin\theta) = I_C(R_C + R_L + jX_C) \quad (16)$$

Comparing (13) with (15) and (14) with (16), it can be found that the system which has two orthogonal transmitter coils is totally equivalent to the system with one transmitter coil which is coaxial with the receiver. (17) shows the correspondence of the parameters in these four equations.

$$\begin{cases} \sqrt{U_A^2 + U_B^2} \cdot \sin\left(\theta + \tan^{-1}\frac{U_B}{U_A}\right) = K_1 U_i \\ I_C = K_2 I_2 \\ M_{BC}\cos\theta + M_{AC}\sin\theta = K_3 M \\ I = K_4 I_1 \\ R + jX = K_5(R_1 + jX_1) \\ R_C + R_L + jX_C = K_6(R_2 + R_l + jX_2) \end{cases} \quad (17)$$

where $K_1, K_2, K_3, K_4, K_5, K_6$ are constants.

From equation (11), the change of $P_{in}$ can be used to identify the presence of the receiver, but it shows no special parameter to identify its type. To find the differences between the normal receiver coil and the metal object, the expressions of $U_i$ and $I_1$ should be gotten first. Rearranging equation (13) and (14) leads to.

$$I_1 = \frac{U_i + j\omega M I_2}{R_1 + jX_1} \quad (18)$$
$$I_2 = \frac{j\omega M I_1}{R_2 + R_l + jX_2} \quad (19)$$

During normal operating conditions, $\omega^2 = 1/L_1 C_1 = 1/L_2 C_2$. Assume that $Z_2 = R_2 + R_l + jX_2$ and substitutes (19) into (18), the expression of $I_1$ can be solved as follows.

$$I_1 = \frac{U_i}{R_1 + \frac{\omega^2 M^2}{Z_2}} \quad (20)$$

From equation (20), what makes the differences between the receiver coil and the metal is the impedance of the receiver $Z_2$ and the mutual inductance $M$. The differences between the two models are shown in Fig. 5. As illustrated in Fig. 5, the equivalent circuit of a metal object is regarded as an R-L model. $Z_{coil} = R_2 + R_L$ when the receiver is a coil, and $Z_{metal} = R_m + j\omega L_m$ when the receiver is a metal object. Moreover, the mutual inductance between the transmitter and the receiver is also different in two models.

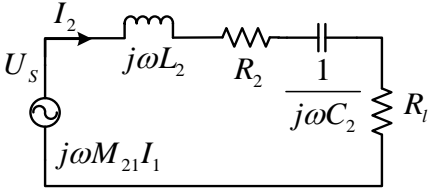

Fig. 5 Equivalent circuit modes of the receiver: (a) coil; (b) metal.

### B. Effect of Metal Object

In order to distinguish whether the receiver is the metal or the coil, the parameters in transmitter part which are mentioned before need to be analyzed in detail. For most metal objects like metal box, copper layers in the PCB board, etc, the portion most affected by the magnetic field is the plane closest to the center of the transmitter coil when the metal objects enter the charging area. Regarding the metal foreign object as a metal plate, the detailed analysis is shown as follows.

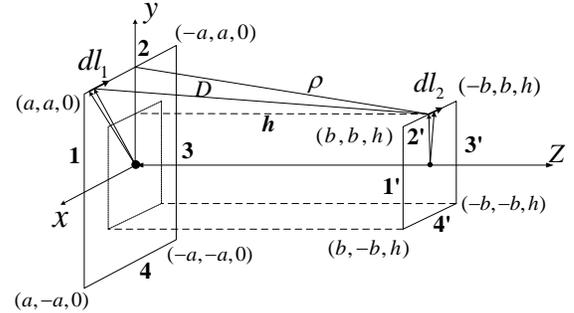

Fig. 6 Structure and position relationship between the transmitter and the receiver in the simplified model.

1) Mutual inductance analysis

The mutual inductance is a constant physic parameter between the transmitter and the receiver, once their shape and relative position are given. Fig. 6 shows the position relationship between the transmitter and the receiver. The transmitter and the receiver are put in a coordinate system with a coaxial structure. $dl_1$ and $dl_2$ are the line elements of the transmitter coil and the receiver, respectively. $D$ is the distance from $dl_1$ to $dl_2$, and $\rho$ is the vertical distance from $dl_2$ to the line where $dl_1$ lies. The distance between the two coils is $h$.

The transmitter coil is considered to be the four current-carrying wires which constituting a square coil. The current in each wire is continuous, so the magnetic flux generated by those wires is continuous too. To calculate the mutual inductance, a new physic quantity is introduced--the magnetic vector potential $\vec{A}$. The magnetic vector potential generated by the transmitter coil in an arbitrary point in the space is shown as follows.

$$A = \frac{\mu_0 \bar{I}}{4\pi} \oint \frac{dl}{D} \quad (21)$$

When the receiver is a coil, the total magnetic flux $\Phi_{coil}$ through it can be expressed as follows.

$$\Phi_{coil} = \oint A \cdot dl_2 \quad (22)$$

And the mutual inductance between two coils is equal to the total magnetic flux through the receiver divided by the current in transmitter coil. The approximate calculation of $M_{coil}$ is shown as follows.

$$M_{coil} = \frac{\Phi_{coil}}{I} = \frac{4\mu_0 b}{\pi} \ln\left(\frac{\sqrt{(a+b)^2 + h^2}}{(a-b)^2 + h^2}\right) \quad (23)$$

When the receiver is not a coil but a metal object with a similar shape, the Microelement Method is used to calculate the mutual inductance between the transmitter coil and the metal plate. The metal plate is seen as a stack of square coils with the radius from 0 to $b$. The derivation of magnetic flux through the metal plate is shown as follows.

$$\Phi_{metal} = \int_0^b \Phi_{coil} db \quad (24)$$

According to equation (24), the mutual inductance between the metal plate and transmitter coil can be calculated as follows.

$$M_{metal} = \frac{4\mu_0}{\pi}[ab + ah \cdot \text{atan}\left(\frac{a-b}{h}\right) - ah \cdot \text{atan}\left(\frac{a+b}{h}\right) + \frac{a^2(\#1)}{4}$$
$$- \frac{a^2(\#2)}{4} - \frac{b^2(\#1)}{4} + \frac{b^2(\#2)}{4} - \frac{h^2(\#1)}{4} + \frac{h^2(\#2)}{4}] \quad (25)$$

where
$$\#1 = \ln(a^2 - 2ab + b^2 + h^2), \#2 = \ln(a^2 + 2ab + b^2 + h^2).$$

In order to find the specific relationship between $M_{metal}$ and $M_{coil}$, the length of the transmitter coil, the length of the receiver and the transmission distance are set artificially. Fig. 7(a) shows the curves of $M_{metal}$ and $M_{coil}$ versus the length of the transmitter coil $a$ when $b = 0.1$ m, and Fig. 7(b) shows the curves of $M_{metal}$ and $M_{coil}$ versus the length of the receiver $b$ when $a = 0.164$ m. The solid lines represent the mutual inductance between coils and the dotted lines represent the mutual inductance between the metal and the coil. Red, blue, green and yellow lines indicated different distance conditions, 0.1 m, 0.2 m, 0.4 m and 0.8 m, respectively. From Fig. 7(a), it's obvious that $M_{coil}$ is much larger than $M_{metal}$ when the size of the coils and metal plate are smaller than 0.5 m. The $M_{coil}$ and $M_{metal}$ achieved the maximum value when the value of $a$ and $b$ are both near to the value of distance $h$. Fig. 7(b) indicate that the larger receiver the larger mutual inductance. Both Fig. 7(a) and (b) show that the mutual inductance is inversely proportional to the distance between transmitter and receiver.

2) Analysis of equivalent impedance of metal plate

The other parameter used to identify the receiver is the equivalent impedance $Z_2$. The value of $Z_{coil}$ is mainly decided by the magnitude of the load $R_L$. And the magnitude of $R_L$ is limited by the magnitude of the output power of the WPT system.

As for $Z_{metal}$, $R_m$ and $L_m$ are both equivalent circuit parameters of a metal plate. $R_m$ is associated with the eddy current loss in the metal object. The origin of $L_m$ is the induced current in the metal object. According to [8], $R_m$ and $L_m$ are calculated as follows.

$$\begin{cases} R_m = \omega\pi\mu_0 \int_0^\infty \Phi_i(k)e^{-2kd}T(k)dk \\ L_m = \pi\mu_0 \int_0^\infty \Phi_r(k)e^{-2kd}T(k)dk \end{cases} \quad (26)$$

where $k$ is the integration variable which represents a spatial frequency, and $d$ is the distance between the transmitter coil and metal plate. $\phi(k)$ is a complex parameter depending on the metal properties and the frequency of the current, and it is defined as follows.

$$\Phi(k) = \frac{\sqrt{k^2 + j\omega\sigma\mu_0\mu_r} - k\mu_r}{\sqrt{k^2 + j\omega\sigma\mu_0\mu_r} + k\mu_r} \quad (27)$$

where $\sigma$ is the conductivity of the metal, $\mu_0$, $\mu_r$ are the permeability and the relative magnetic permeability of a metal, respectively. $\phi_r$ and $\phi_i$ are the real and imaginary parts of $\phi(k)$, respectively (i.e. $\phi = \phi_r + j\phi_i$). Besides, $T(k)$ is a geometric function defined as

$$T(k) = [NaJ_1(ka)]^2 \quad (28)$$

with $J_1$ being the Bessel functions of the first kind of order 1. And $N$ is the turns of the transmitter coil. $a$ is a half-length of the transmitter coil.

From equation (26)-(28), $R_m$ and $L_m$ can be calculated numerically if the size and the material properties of the metal and the transmission distance are already known. Table I shows several properties of different metallic material.

TABLE I
METAL PROPERTIES

| Material | $\sigma_1 (\Omega m)^{-1}$ | $\mu_r$ |
|---|---|---|
| Cuprum | $5.88 \times 10^7$ | 1 |
| Aluminum | $3.44 \times 10^7$ | 1 |
| Ferrum | $1.00 \times 10^7$ | 200~400 |

Assume that $d = 0.20$ m, frequency $f = 20$ kHz, the curves of $R_m$ and $L_m$ versus the length of the metal plate are shown in Fig. 8. In Fig. 8, both $R_m$ and $L_m$ increase with the increment of the length of the metal plate. It can be seen that the ferromagnetic and the Non-ferromagnetic material show a great difference in equivalent resistance while their self-inductance are almost the same.

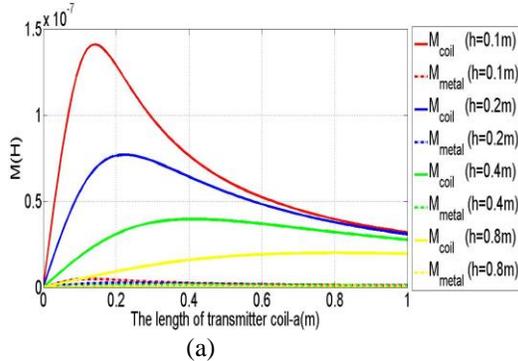

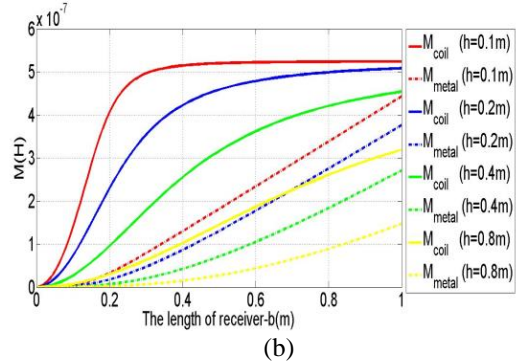

(a) (b)

Fig. 7 the value of mutual inductances versus the length of transmitter and receiver: (a) transmitter-a; (b) receiver-b.

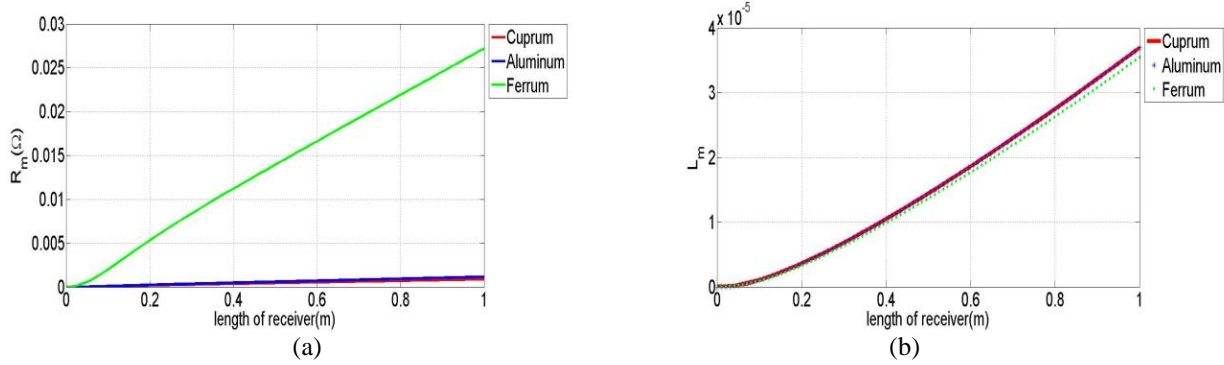

Fig. 8 The value of metal plate equivalent resistance and inductance versus the length of metal plate: (a) resistance $R_m$; (b) inductance $L_m$.

3) Method for MOD

After the detailed parametric analysis, this part takes the parameters from Fig. 7 and Fig 8 into equation (11) and (20) to draw the load impedance and the input power curves of different receivers. According to Fig.7, this part divides the concrete analysis into two cases. One is when $a$=0.164 m, $b$=1 m, and $M_{coil}/M_{metal} = 1.3$. The other case is when $a$=0.164 m, $b$=0.1 m, and $M_{coil}/M_{metal} = 29$. Ferromagnetic material has the largest resistance in proposed metal, so only the load impedance and input power characteristics of Ferrum are analyzed. The following Figs show the relationship of the circuit parameters in transmitter coils. $I_{TX}$ and $U_{TX}$ are the current and voltage in the transmitter coil, respectively. $P_{input}$ is the input power of the system.

It can be seen from the curves in Fig. 9 and Fig. 10, the refracted impedance of the metal object in the primary side is substantially smaller than that of the receiver coil. But according to the coupled circuit model and equation (20), the equivalent impedance of the metal is much larger than the coil with the load. When $I_{TX}$ increases, the increment of $U_{TX}$ and $P_{input}$ is relatively less if the receiver is a metal object. The method for MOD can be obtained from the illustrated curves. The curves shown in Fig. 9 are based on the condition when the receiver is much larger than the transmitter coil. And as the length of metal increases, $R_m$ grows faster than $L_m$. Therefore, by setting the load impedance and input power characteristic curves of the 'largest' metal object as the threshold value, a reliable function of metal object detection can be achieved. 'Largest' means all the magnetic flux generated by the WPT system is absorbed by the metal. In practical application, the load impedance and input power characteristic curves of this 'largest' metal can be obtained by statistics and regression algorithm.

Simulation calculation has also considered the critical case when the impedance of coil and metal are exactly the same. Since the inductive impedance of the metal plate is much larger than the equivalent resistance of the eddy current loss at high frequencies, the input power curve corresponding to the metal object will still below the input power curve corresponding to the coil.

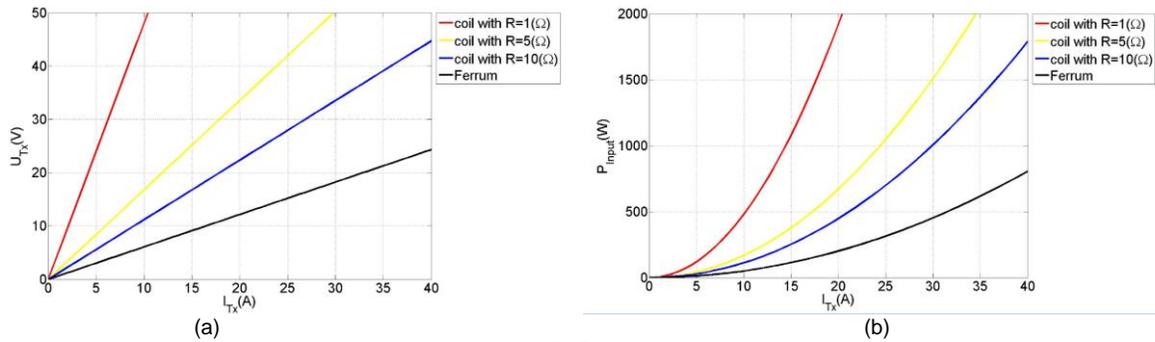

Fig. 9 (a) relationship between $I_{TX}$ and $U_{TX}$ when $M_{coil}/M_{metal} = 1.3$; (b) Relationship between $I_{TX}$ and $P_{Input}$ when $M_{coil}/M_{metal} = 1.3$.

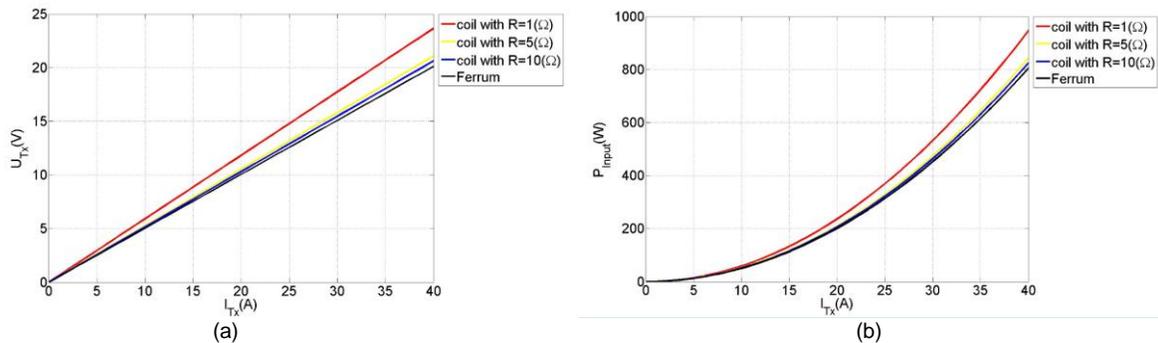

Fig. 10 (a) Relationship between $I_{TX}$ and $U_{TX}$ when $M_{coil}/M_{metal} = 29$; (b) Relationship between $I_{TX}$ and $P_{input}$ when $M_1/M_2 = 29$.

## IV. EXPERIMENTAL VERIFICATION

### A. Hardware Setup

In order to verify the proposed detection method, a prototype of the WPT system is implemented as Fig. 11 shows. The prototype is composed of a controller which equipped with a floating-point digital signal processor (DSP.TMS320F28335) and a field programmable gate array (FPGA.EP2C8J144C8N), an AC/DC converter with two independent channel 12 V output, two independent controllable DC/AC converters, two independent resonance tanks including series resonant capacitors and transmitter coils. Coil A labeled red and coil B labeled yellow, each coil is a square with the length of side 328 mm, each coil has three turns of Litz wire with radius 2.2 mm. The receiver coil which labeled green is connected with a series resonant capacitor and a resistance load. The receiver coil has the same structure of coil A and coil B. The inductance of each transmitter coil and the receiver coil is designed as 10 $\mu$H, the capacitance of each series resonant capacitor for transmitter and receiver are designed as 6.6 $\mu$H. The frequency of AC voltage generated by DC/AC converters is designed as 20 kHz, DSP is in charge of mathematical calculation and FPGA is in charge of generating PWM signals for switches.

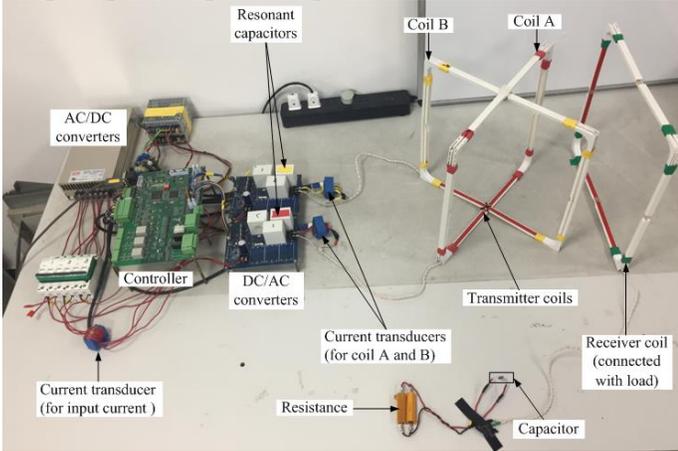

Fig. 11 Implementation of a wireless power transfer system.

### B. Verification of MOD Method

The prototype is working at the frequency of 20 kHz. The metal plate is a square with the size of 200 mm$^2$. The materials of the tested metal plates are Ferrum (Fe), Aluminum (Al) and Cuprum (Cu). They will be placed in the direction, where the total magnetic flux points to, with their plane facing the center of the transmitter coils. And the distance from the center of transmitter coils and the center of the receiver is 20 cm. The current in the transmitter coils is set as a constant—10 A. Hence, the magnitude of the magnetic generated by the prototype will always be the same. The waveforms in different conditions are shown in Fig. 12-13. The waveforms from top to the bottom are the output AC voltage on the resonant circuit, the input DC current of the total power supply, the current in coil A and the current in current B, respectively. The waveforms of different metal plates are almost the same as the waveforms in Fig. 13, except the RMS of each waveform. The test results can be found in Table II.

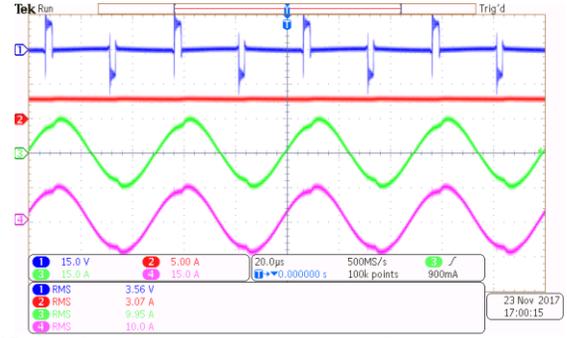

Fig. 12 Waveforms when the receiver is a coil with resonant capacitor and a resistive load of 4.5 Ω.

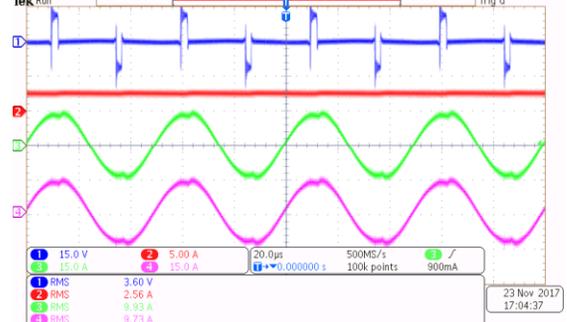

Fig. 13 Waveforms when the receiver is Ferrum plate with the length of 200 mm.

TABLE II
TEST RESULTS OF DIFFERENT RECEIVERS.

|  | $U_{AC}$ | $I_{DC}$ | $U_{coilA}$ | $U_{coilB}$ |
|---|---|---|---|---|
| Coil(1.5Ω) | 3.83V | 3.34A | 9.83A | 10.0A |
| Coil(4.5Ω) | 3.56V | 3.07A | 9.95A | 10.0A |
| Cuprum | 3.60V | 2.58A | 9.83A | 9.73A |
| Aluminum | 3.53V | 2.55A | 9.74A | 9.77A |
| Ferrum | 3.60V | 2.56A | 9.93A | 9.73A |

The input DC voltage is constant at 12 V during the experiment. The input DC current represents the input power of the prototype. From Table II, when the current in the transmitter coil is the same, the coil load has a higher voltage on the transmitter coil which is coincident with Fig. 9 and Fig. 10. And the coil with 4.5 Ω resistance is like the critical state. Whether the receiver is a coil or a metal object, the voltage and current of the transmitter coil are almost the same. However, the metal object is still distinguishable in this condition in this case, because the input power is higher in the coil's case.

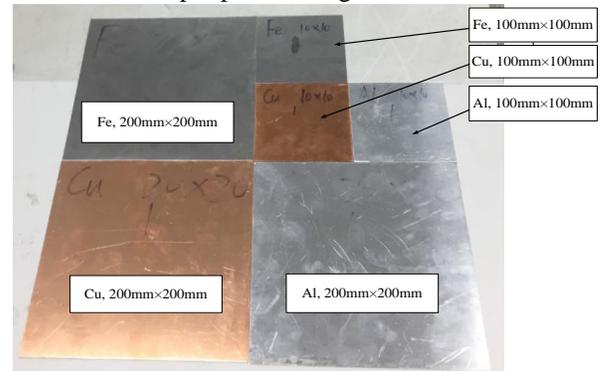

Fig. 14 Metal plates used to test the method for MOD.

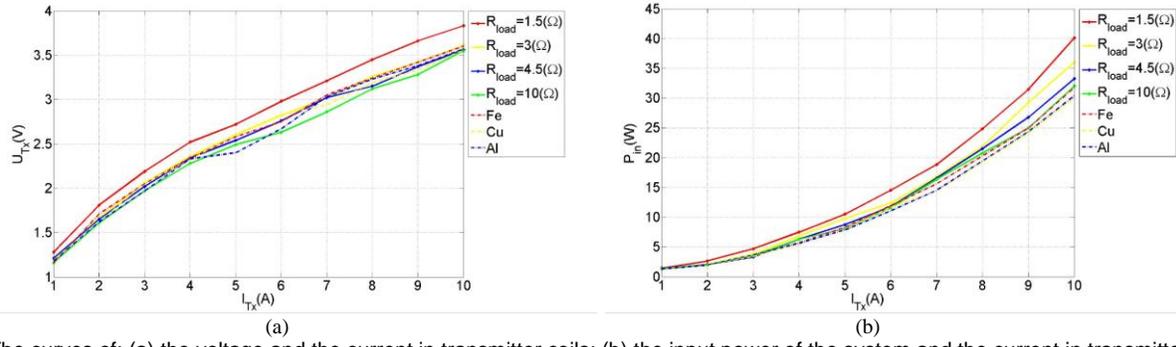
Fig. 15 The curves of: (a) the voltage and the current in transmitter coils; (b) the input power of the system and the current in transmitter coils.

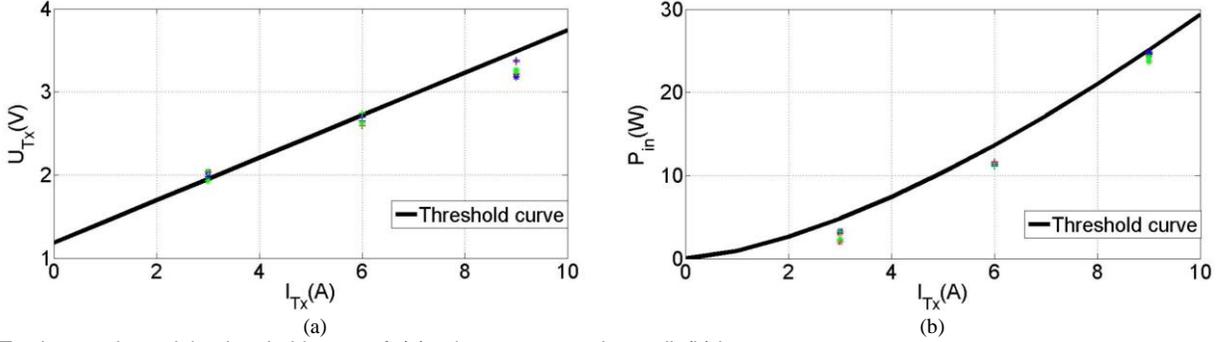
Fig. 16 Testing results and the threshold curve of: (a) voltage on transmitter coil; (b) input power.

Next, the load impedance and input power curves are drawn from the statistics data, and the threshold curves are obtained by the regression algorithm. When different metal objects enter the charging area, the voltage on transmitter coils and the input power of the system are collected and compared with the threshold curves.

As Fig. 14 shows, three different material of metals which are commonly used in the home application are chosen to test. Currents in transmitter coils and input DC current are sampled by current transformers. The information collected from the transmitter side will be sent to the DSP where the code of the method for MOD lies in. The current in each transmitter coil is controlled by DPS. The current in the coil is set to 0-10 A sequentially, and the corresponding coil voltage and the system input power are recorded. The results of the information collecting are shown in Fig. 15 in the form of curves. The curves of metal are experimentally obtained with the larger metal plate shown in Fig. 14.

In Fig. 15 (a), when the load resistance is less than 4.5 Ω, the corresponding impedance characteristic curves of different load are above that of the metal. However, in the extreme case where the load is 10 Ω, the impedance characteristic of the load becomes hard to discern. The input power curves shown in Fig. 15 (b) are another guarantee for the proposed detection method. It can be seen that even the load is 10 Ω, the corresponding power curve is still higher than metal.

After the statistics, the threshold curves are set at the intersection of the metal curves and the coil curves by regression analysis of those experimental data. The threshold curve of U-I is obtained by linear regression and that of P-I is obtained by power series regression. After that, each metal plate in Fig. 14 will be tested when the current in transmitter coils are 3A, 6A and 9A, respectively. The testing results are recorded and shown in Fig. 16. The points marked red, blue and green represent the metal of Fe, Cu and Al, respectively. And different points '*' and '+' represent the size of the metal object is 100 mm$^2$ and 200 mm$^2$, respectively. In Fig. 16, the point below the threshold curves will be considered as metal objects. From the experimental results, the metal can be accurately detected when the current in transmitter coil is greater than 3 A. Because when the coil current is too small, the Gaussian noise can be relatively larger, that may cause detection failed. Through the analysis and verification of the experimental part, it can be concluded that the detection method proposed in this paper can effectively detect the existence of metal objects, regardless of the size of the metal object.

V. CONCLUSION

A MOD method based on the load impedance and input power characteristic is proposed in this paper to protect high dimensional WPT system from the influence of the surrounding metal objects. Based on the equivalent models of the WPT system, the calculation of mutual inductance and the discussion of the influence caused by metal and receiver coil show the essential difference between metal and coil. By using the current and voltage in transmitter coils and the input power of the system, different curves are drawn in U-I and P-I plots which indicates the load impedance and input power characteristic of different receivers. According to the simulation calculation, all the curves of metal are under the curves of the coil. By this characteristic, the proposed method is validated in the prototype. And the way to set the threshold curve is described in the experiment. High-dimensional WPT system using this method will be able to ensure the safe operation in no-communication condition. The system can effectively shut down or standby by artificial settings when the metal is detected. The proposed method also can be developed for a three crossed-loop coils WPT system with higher degrees of freedom.